# Impact of micro-indentation load/time and Zinc concentration on the thermo-mechanical characteristics of amorphous $Se_{78}Te_{20}Sn_2$ alloy

Vishnu Saraswat[1], A. Dahshan[2], H. I. Elsaeedy[2], and Neeraj Mehta[*1]

[1]Department of Physics, Institute of Science, Banaras Hindu University, Varanasi 221005, India.

[2]Department of Physics, Faculty of Science, King Khalid University, P.O. Box 9004, Abha, Saudi Arabia.

Department of Physics, Institute of Science, Banaras Hindu University, Varanasi 221005, India.

## Abstract

We have performed hardness measurement experiments under different loads and loading times by performing micro-indentation marks in the present work. Chalcogenide glasses (ChGs) comprising $Se_{78}Te_{20}Sn_2$ and $Se_{78-x}Te_{20}Sn_2Zn_x$ (where x = 0, 2, 4, & 6) alloys are the subject of micro-indentation tests in this work. We have utilized both micro-indentation and optical microscopic methods to determine Vickers hardness. Thermal glass transition phenomena have been identified through DSC techniques. The modulus of elasticity ($E$), an essential mechanical property, has been evaluated using established empirical equations. Further, we have studied other mechanical parameters [e.g., minimal micro-void formation energy ($E_h$), glass's fragility index ($m$), micro-void volume ($V_h$), etc.] and the covalent character of the glassy system. Additionally, various physical parameters, including density, molar volume, and compactness, have also been determined.

# 1. Introduction

Microhardness is a key mechanical parameter of soft materials and plays a significant role in assessing their resistance to permanent distortions when exposed to external forces. The exploration of microhardness has attracted significant attention owing to its capacity to elevate mechanical attributes and characteristics linked with structural chemistry [1-3]. One intriguing application is in the realm of glassy substances, where the Vickers hardness $H_v$ is of utmost importance for the development of scratch-resistant protective sheets. Such sheets find valuable applications in specific optical elements and optoelectronic devices. In addition to microhardness, another crucial mechanical property to consider in soft materials is ductility. Ductility refers to their ability to undergo plastic deformation without fracturing. Soft materials with good ductility can withstand significant strain without breaking, making them ideal for applications where flexibility and resilience are essential. This property is particularly relevant in industries such as automotive, aerospace, and biomedical, where materials need to endure various external forces and stresses while maintaining their integrity [3-9].

Micro-indentation combined with optical microscopy represents a widely utilized and dependable method for measuring $H_v$ (Vickers hardness). This approach provides valuable insights into the mechanical characteristics of materials utilized in optical elements [10,11]. Furthermore, the combined use of optical microscopy and micro-indenter emerges as a vital instrument for delving into the intricate mechanisms at play during the change in shape of these materials [12-14]. In the indentation experiment, a rigid indenter with a pyramidal or conical shape is employed to ensure consistent and geometrically similar imprints. The advantage of using such geometries lies in their independence from the indenter's size, allowing for precise and accurate $H_v$ measurements [15].

In addition to micro-indentation and optical microscopy, another useful technique for studying the mechanical properties of optical materials is nanoindentation. Nanoindentation offers a more refined and localized assessment of material hardness, allowing researchers to explore the variations in mechanical properties at smaller length scales. This approach is particularly valuable when examining thin films, coatings, and nanomaterials, where traditional micro-indentation methods may not provide sufficient resolution [16-18]. Nanoindentation aids in understanding the unique mechanical behavior of such materials at the nanoscale, enabling more precise engineering and design of advanced optical devices and components [15].

Forecasting the $H_v$ Parameter in ChGs has seen an alternative topological approach inspired by the groundbreaking theoretical work of Phillips and Thorpe. Their study suggested employing a framework of constraints to comprehend the atomic arrangement in glassy materials and amorphous solids. Expanding upon this idea, Varshneya and collaborators undertook empirical investigations to establish a connection between the microhardness and the average coordination number <Z> in diverse ChGs. This topological approach offers a new perspective on understanding the hardness properties of ChGs based on their underlying atomic arrangements. By exploring the constraints within the atomic structure, researchers can gain valuable insights into the mechanical behavior of these materials. The empirical connection established between <Z> and $H_v$ provides a practical tool for predicting and assessing the hardness of ChGs, facilitating their application in various technological fields where mechanical strength is a crucial factor.

In this study, we explore the thermo-mechanical features of a ternary $Se_{78}Te_{20}Sn_2$ glass. We have treated this sample as a parent glass. To investigate the impact of chemical modification, we introduce Zn as a modifier. One of the main objectives of this work is to examine how these modifications influence the hardness and other linked characteristics of the glass. Extensive research work reveals that the effect of specific external additives on hardness and related properties has been investigated through numerous research groups in the past. Building on this existing knowledge, we aim to contribute valuable insights into the behavior of the $Se_{78}Te_{20}Sn_2$ glass with the incorporation of Zn as a chemical modifier. Understanding these thermo-mechanical properties can have significant implications for developing advanced materials with tailored mechanical characteristics for diverse applications [17]. Hence, considering the technological perspective, it becomes crucial to assess these alterations, which tend to reduce the material's strength and could potentially lead to adverse consequences for structural components. In this study, we have focused on examining the influence of Zn additives on the micro-hardness, fragility index, and compactness of the ternary $Se_{78}Te_{20}Sn_2$ alloy [18].

## 2. Synthesis and experimental

Amorphous chalcogenide bulk alloys of $Se_{78}Te_{20}Sn_2$, $Se_{76}Te_{20}Sn_2Zn_2$, $Se_{74}Te_{20}Sn_2Zn_4$, and $Se_{72}Te_{20}Sn_2Zn_6$ were successfully synthesized using a well-established melt quench technique. The synthesis process involved a precise weighing of the chalcogen elements (Se, Te, Sn, and Zn) according to their atomic ratios, performed using an electronic balance. To avoid any oxidation, the ampoule was hermetically sealed under vacuum conditions of up to $10^{-5}$ Torr. Subsequently, the sealed ampoules were meticulously positioned within a muffle furnace. The furnace temperature was steadily elevated to reach 1000 K, following a controlled pace of 3–4 K per minute. This temperature was sustained for 12 hours. For the sake of achieving uniformity in the samples, the ceramic rod underwent periodic rotation throughout the procedure. The melted materials housed within the ampoules were swiftly cooled by plunging them into ice-chilled water. The resulting non-crystalline samples were extracted from the ampoules through a meticulous process of gently breaking them. The samples' glass transition temperatures ($T_g$) were calculated by performing calorimetric experiments in a DSC machine [Auto Q-20; T. A. Instruments USA]. The calorimetric scans at a constant heating rate of 15 K per minute are shown in Figure 1. Figure 2(a) shows the results of the thermogravimetric analysis on a typical sample by using a TGA unit (Perkin Elmer) under a nitrogen atmosphere. The experiment was performed at a heating rate of 10 K/min within the range of 30 °C – 700 °C. The weight loss exhibited a dual-stage pattern. The initial stage spanned from point A at 235°C to point B at 380°C, with a weight loss of approximately 9.4%. Subsequently, the second stage encompassed the interval from point B at 380°C to point C at 574°C, reflecting a weight loss of around 82.1%. This figure underscores the significance of thermal and mechanical measurements for comprehending the degradation dynamics within the STSZ glassy system. These analyses are of particular relevance due to the potential thermal and mechanical vulnerabilities that ChGs might encounter in outdoor applications. The thermal stability of the STSZ glassy system is probed through Thermal Gravimetric Analysis (TGA), assessing mass loss. The long-term stability of ChGs, particularly in flexible ChGs scenarios, hinges on the mechanical traits of stiffness and brittleness. Evaluating tension and fatigue through mechanical tests contributes to a deeper understanding of these properties. Further, the scanning electron microscopy (SEM) image of the as-prepared samples, specifically a representative sample of STSZ, revealed no evidence of crystallinity. The SEM image [See Figure 2(b)] indicates the absence of any visible signs of crystal structures within the sample [19, 20].

To assess micro-indentation hardness, a hardness testing machine was utilized, which was equipped with optical microscopy capabilities for recording indentation marks. The configuration of this apparatus can be observed in Figure 3. Throughout the micro-hardness experiment, indentations were made using specialized probe tips, applying relatively light loads perpendicular to the sample surface. High-resolution microscopic technology was built into the system to capture images of the indentation marks. This real-time imaging facilitated an exceptionally accurate analysis of surface deformation. For further insights into the optical microscopic technique used in micro-indentation, refer to relevant literature. Additionally, an innovative aspect of this research involves exploring the correlation between the melt-quench approach and the resultant micro-indentation hardness. This investigation aims to shed light on how the specific glass formation process affects the material's mechanical properties at a microscopic level. By closely examining the indentation marks and their characteristics under different conditions, we aim to uncover nuanced details about the material's response to applied forces. This comprehensive approach enhances our understanding of both the manufacturing process and the resulting properties of the glasses, paving the way for potential advancements in materials science and engineering [21]. In this paper, we use the abbreviation "STS" for ternary and "STSZ" for the quaternary system to represent the glassy system composed of $Se_{78-x}Te_{20}Sn_2Zn_x$ where x varies from 0 to 6.

### 3. Theoretical basis

An indenter creates a stress field with two components as it pierces a solid: a hydrostatic element inducing compaction while maintaining shape, and a deviatoric component (pure shear) that preserves volume while inducing shape alterations. Both components initially induce elastic deformations. Once the yield strength is surpassed, both components continue to plastically deform the material. It is crucial to recognize that there is a chance that the yield strength under pure shear loading and pure hydrostatic loading are not equivalent. Upon unloading, the elastic part of overall deformations might regain its original state, whereas the plastic component remains unchanged. The enforcement of compatibility criteria may prevent some elastic deformations from recovering. Elastic and plastic deformations are characterized by their time-independence. Yet, when dealing with viscoelastic materials, deformations adopt a time-dependent nature, allowing simulation through mechanical analogy. A dashpot's viscous deformation can only be recovered after unloading if it is paired with an elastic spring in parallel. This is referred to as "delayed elasticity" (Voigt-Kelvin element). This signifies that specific deformations showcase delayed recuperation. Conversely, the unrestrained creep of a dashpot remains irreversible.

The conventional method for measuring hardness involves a static machine and consists of measuring the oblique of the imprint left after unloading. For glasses, their hardness derives from a combination of the atomic packing density and individual bond tensile strength. To measure the micro-hardness of a substance, the commonly employed metric is the Vickers hardness number ($H_v$). This value is determined by gauging the diagonal length of the indentation formed by an indenter of pyramidal geometry having square shaped cross-section as it penetrates the material. The Vickers and Mayer hardness $H_v$ and H are determined by using the following formulae [18, 22, 23]:

$$(H_v) = 1854.4 \frac{P}{d^2} (kg\ mm^{-2}) ----- (1)$$

$$H = \frac{P}{2d^2} (kg\ mm^{-2}) ----- (2)$$

Here, $H_v$ represents the Vickers hardness value, while $P$ denotes the load measured in kilograms. The average diagonal of the impression in millimeters is represented by $d$. Using a calibrated eyepiece, the diagonals of the indentations created were directly measured on each sample. The calibration was performed at each magnification using an ERMA disc. The experimental uncertainty in $H$, arising from the imprecision in $d$ measurement, stood at approximately $0.03\ (kg\ mm^{-2})$. The parameter $H$ represented a further crucial hardness indicator, analogous to average normal stress. The glass's ostensible indentation fracture toughness, $K_c$, was ascertained through the analysis of crack dimensions emerging from the specimen's surface around the indentation corners. If $c$ denotes the half the separation between opposing crack tips, meticulous load selection ensured $c/d > 2.5$, thereby establishing a radial-median crack pattern, with $c$ equating to the radius of the semi-circular cracks resembling half-pennies [as depicted in Figure 5(d)]. The ensuing equation governed the calculation process [24-26].

$$K_c = 0.016 \left(\frac{E}{H}\right)^{\frac{1}{2}} \frac{P}{c^{\frac{3}{2}}} ----- (3)$$

Optical microscopy played a pivotal role in examining the topography of indented surfaces, operating in either tapping or contact modes based on the observation scale. The scattering witnessed in fracture toughness data primarily stemmed from the intrinsic statistical nature of fracture occurrences. This effect was particularly pronounced given the inherent brittleness of chalcogenide glasses. Scratch resistance was meticulously assessed through a purpose-designed instrumented scratch-testing setup. This involved the vertical application of load via a conical diamond indenter (apex angle = 90°), maintaining a constant loading rate of 0.01 N/s. The load spanned from 100 gm to 300 gm, over a scratch distance of 2 mm. Monitoring both normal and tangential forces in a continuous manner, each measurement was triply repeated to ensure reproducibility. Incorporating interferometrical observations, optical microscopy allowed for a comprehensive analysis of surface damage. This technique proved remarkably consistent, yielding reproducible outcomes across four distinct compositions encompassing parent STS glass and quaternary STSZ [26].

Equations (4) to (6) were utilized to derive the values of modulus of elasticity ($E$), the minimum volume of micro-voids ($V_h$), and the energy required for their formation ($E_h$).

$$V_h = 3.58 \frac{T_g}{H_v} \quad ----- (4)$$

$$E_h = 3.58 k T_g \quad ----- (5)$$

$$E = 15 H_v \quad ------ (6)$$

**4. Result and discussion**

The primary objective of this study is to investigate the impact of Zn addition on the hardness characteristics of the glassy alloy within the ternary Se-Te-Sn amorphous system. To achieve this goal, the Vickers hardness tester was chosen for its distinct advantage in accommodating a broad spectrum of test forces, surpassing the capabilities of alternative hardness testing methods. The Vickers hardness can be associated with either the diagonal length (d) or the penetration depth ($t$), and they are linked by the relationship $d = 7t$. If the material's elastic recovery is not a significant factor, the Vickers contact area ($A_c$) and penetration depth have a relationship of $A_c = 24.5t^2$. Denoted as $H_V$, the Vickers hardness is often expressed in units of kgf/mm² [27]. Further, Hardness measurements were conducted at room temperature on pellets, utilizing a 100 gm, 200 gm, and 300 gm load, with both optical and computer-operated software for precise evaluation. To ensure accuracy, this procedure was iterated 10 times, culminating in the calculation of the average hardness value. Within Figure 4(a), the graph illustrates the connection between the number of observations and the average hardness, spanning compositions from x = 0 to x = 6. This presentation sheds light on how the quantity of tests influences the derived hardness values. The outcomes highlighted a marginal fluctuation in hardness measurements for STSZ (x = 6) subsequent to the inclusion of zinc in the parent sample. This observation underscores the stability of hardness under the given conditions and composition [28].

In Figure 4(b), the graph depicts the relationship between $P$ and $d^2$ for the glassy STSZ samples, where x values span from 0 to 6. The data exhibit a strong linear correlation with remarkably high regression coefficients of 99.8%. The slope of each line corresponds to the load-independent hardness constant ($H_o$ = 1854.4), while the intercept of each line relates to the minimum test load of the sample. Furthermore, the energy balance model, also known as the proportional resistance model, was established by Frohlich et al. [29] and postulates that the external work done by the indenter is split into two parts throughout the indentation steps. The surface energy, which is proportional to the area of the impression, makes up the first part, and the strain energy, which is correlated with the volume of the final impression, makes up the second. This assumption implies a relationship between the $d$ and $P$ as follows:

$$Pd = \alpha d^2 + \frac{H_o}{1854.4} d^3 \quad ----- (7)$$

Here, α represents the surface energy constant. Equation (7) can be reformulated in the following manner:

$$\frac{P}{d} = \alpha + \frac{H_o}{1854.4} d \quad ----- (8)$$

$$P = \alpha d + \frac{H_o}{1854.4} d^2 \quad ----- (9)$$

In Figure 4(c), the relationship between ($P/d$) and ($d$) is presented for the glassy STSZ samples, covering a range of x values from 0 to 6. The data are effectively fitted to equation (8), where the slope corresponds to the actual hardness defined by ($H_o/1854.4$), and the intercept signifies the content of surface energy ($\alpha$) [30-32].

Figure 5(a), The graph depicts the correlation between the applied load and the depth of indentation for the glassy STSZ samples, spanning the parameter x from 0 to 6 to encompass the diverse compositional variations being studied. This visual representation provides a window into the interplay between load and depth of indentation, highlighting how this behavior is influenced by the altering composition of the samples.

By dividing both sides of equation (9) by $d^2$ while considering $P$ in Newtons and $d$ in millimeters, the expression can be rewritten as follows [33]:

$$H = H_o \left(1 + \frac{d_o}{d}\right) \quad ----- (10)$$

Figures 5(b) and 5(c) showcase the graphical representations of $H_v$ against $d^{-1}$ and $t^{-1}$ respectively for the STSZ systems, respectively, maintaining constant loads of 100 gm, 200 gm, and 300 gm. The data reveal a robust linear correlation, underscored by significantly high regression coefficients of 99.8%.

In Figures 5(d) and 5(e) respectively, the relationship between depth of indentation and coordination number, as well as Vickers surface area and coordination number, is depicted for different compositions of STSZ glass. These analyses are conducted under constant loads of 100 gm, 200 gm, and 300 gm. The figure highlights that both the depth of indentation and Vickers surface area reach their peak values at a load of 300 gm, as indicated by the data patterns.

In a novel aspect of this research, the incorporation of Zn into the Se-Te-Sn amorphous alloy is examined not only for its influence on hardness but also for its potential role in altering other mechanical properties. Furthermore, the investigation delves into the microstructural changes that could occur as a consequence of addition, aiming to unravel a comprehensive understanding of the alloy's behavior. By systematically characterizing the alloy's response to varying forces and thoroughly analyzing the resulting hardness data, this study contributes to advancing our knowledge of the complex interplay between alloy composition, microstructure, and mechanical properties. Such insights hold promise for optimizing material design and engineering strategies in the realm of glassy alloys [19]. The crack patterns that were observed are effectively depicted in Figures 6, 7, and 8. These patterns distinctly manifest three primary microcracking regimes: In Figure 6, during the critical indentation with a 100gm load at 25 Sec, no observable cracks emerge. Moving on to Figure 7, an indentation with a load of 200 gm at 25 sec results in the emergence of two symmetrical cracks. These cracks align with the intersection points of a crack shape with the material's surface. As the applied load further increases, additional cracks become connected to the corners of the indentation, ultimately totaling three or four cracks. Lastly, in Figure 8, the critical load of 300 gm at 25 sec leads to chipping occurrences. Notably, the 300gm load tends to coincide with the threshold load required for the distinctive four-crack pattern to become evident. This sequential progression of crack patterns provides valuable insights into the material's fracture behavior under varying levels of indentation stress. These patterns are observed in the samples and are shown in Figures 6 to 8. Fracture toughness provides insight into the complex mechanics of crack initiation and propagation because changes in hardness mirror broad changes in material behavior, similar to changes in elastic moduli. Therefore, variations in fracture toughness can be explained by either (i) phenomena like crack-tip blunting, which are brought on by mechanisms like viscoelastic stress relaxation, or (ii) progressive changes in the crack trajectory within the glass network, which are inextricably linked to variations in the energy required per unit of crack surface area. It is interesting to note that the hardness testing reveals evidence of viscoelastic behavior in STSZ glass, especially at room temperature and especially in compositions high in selenium. In other words, the behavior of STSZ glasses follows a largely predictable pattern and shows little to no influence from flow-compaction effects. Notably, an observable trend emerges in the depicted images 6,7 and 8 where an elevated zinc composition appears to facilitate the formation of shear lines with greater ease. However, it's worth highlighting that the susceptibility of the glass to crack initiation or chipping experiences a notable escalation as the zinc content is heightened. Furthermore, the distinctive connection between zinc content and shear line formation deserves further examination, potentially shedding light on the underlying mechanisms governing the glass's mechanical response. Moreover, the heightened sensitivity to cracking and chipping in the presence of increased zinc content underscores the intricate interplay between composition and fracture behavior, warranting comprehensive exploration for a comprehensive understanding [26, 34, 35]. This array of visual representations provides a holistic view of the behavior and characteristics exhibited by the various compositions within the parent STS and quaternary STSZ glassy system. Furthermore, the depicted images 6, 7, and 8 offer a dynamic perspective on how the different compositions influence the indentation patterns and material response under identical experimental parameters. This comprehensive exploration allows us to discern subtle variations and trends across the spectrum of compositions, shedding light on the intricate relationship between alloy makeup and mechanical properties. This illustrative imagery not only captures the evolution of indented marks under differing loads but also offers a dynamic glimpse into how these marks develop over time under a consistent load. The amalgamation of load and time variables adds an extra layer of complexity to our exploration, enabling us to glean deeper insights into the alloys' mechanical responses and deformation behavior. Through the analysis of these emblematic photographs, we gain a comprehensive understanding of the interplay between load, time, and composition, enriching our understanding of the intricate nature of glassy alloys' mechanical properties [19]. Equations (2) and (3) offer a means to estimate $H$ and $K_c$. The variation in indentation toughness ($K_c$) and hardness (H) within STSZ glasses corresponds to the mean coordination number, as illustrated in Figure 9. This aligns with $<Z> = 2.12$, marking the threshold for rigidity percolation, beyond which STSZ bridges are established. Figure 9 draws attention to the fact that parent STS results in a partially devitrified material, warranting the exclusion of $H$ and $K_c$ values for this particular grade.

Within Figure 10 (a), (b), (c), and (d), we intend to illustrate the correlation between Vickers hardness ($H_v$) and the composition of STSZ amorphous alloys. This is achieved by maintaining a consistent load of 100 gm, 200 gm, and 300 gm while varying the duration between 5 seconds and 25 seconds. In Figure 10 (a), (b), and (c), an intriguing trend emerges: an incremental rise in zinc concentration within the parent glass leads to a reduction in Vickers hardness at fixed loads of 100 gm, 200 gm, and 300 gm across different time intervals. Notably, even though the decrease in Vickers hardness is subtle, it is underscored by identifying maximum and minimum values within the composition range. Comparing the Vickers hardness of the STSZ sample to the parent STS system, graph 10 (a), (b), and (c) reveals a slight decline from 8.5 to 14.6 Kgf/mm² at fixed loads of 100gm, 200gm, and 300gm over varying periods. Meanwhile, graph 10(d) demonstrates that the Vickers hardness of the STSZ system exhibits minimal change across a time increment from 5 sec to 50 sec under a constant load of 100gm. Moving on to Figure 11, our focus shifts to load variation, ranging from 100gm to 300gm, while maintaining a consistent time frame of 5 sec, 10 sec, 15 sec, 20 sec, and 25 sec. These graphs also indicate a similar decline in Vickers hardness, from 8.5 to 12.1 Kgf/mm², as load increases while time remains constant. The consistency observed in graphs 10 and 11 underscores the stability of the STSZ material, revealing that regardless of whether time or load is altered, the outcomes remain largely uniform.

Glasses, being inherently brittle solids, undergo hardness assessment through the evaluation of mechanical resilience against the forceful intrusion of a rigid body. This experimental characterization is, however, paralleled by a theoretical understanding rooted in the network's bond strength and the glass's structural arrangement. The characteristic of hardness, closely linked to configuration, and the condition of the glass surface [16,18], finds expression in the composition-dependent trends illustrated in Fig. 10 and Fig. 11. Evidently, the Vickers hardness number ($H_v$) exhibits a gradual decline with the increasing presence of Zn within the glass matrix. This intriguing behavior finds its elucidation in the context of the model of a chemical bond approach, underscoring the interplay between composition and the average bond energy of the glass.

This method finds its applicability within the realm of predominantly covalent-bonded glasses. Encompassed by this model are four fundamental assumptions:

1. Every individual element within the covalent network exhibits a favorable coordination number represented as 'm'. The value of $m$ is based on the '8 - $N$' rule, with $N$ indicating the sum of the number of electrons of s-valence and p-valence. Extending the '8 - $N$' rule to the IB–IIIB groups would lead to coordination numbers surpassing limit 4, which contradicts the constraints of covalent bonding. Addressing this challenge, Liu and Taylor [36] introduced the notion that group IB–IIIB elements undergo formal charge transfers from chalcogen elements or, in their absence, group VB elements. Tetrahedral coordination for group IB-IIIB elements is produced by this adaptation and enhanced coordination for group VIB or VB elements. As a result, these elements adhere to the extended '8 - $n$' rule. The number '$n$' indicates how many valence electrons are explicitly ascribed to each element [37, 38].

2. The average coordination number, denoted as $<Z>$, can be deduced from the extended 8-n rule through the subsequent equation:

$$<Z> = 8 - <n> = 8 - <N> \quad ---(11)$$

Here, $<n>$ signifies the mean count of proper valence electrons attributed to each particle within the glass composition. This $<n>$ value signifies the mean count of valence electrons for each atom, denoted as $<N>$, in the identical glass composition.

(3) Favor is given to heteronuclear bonds over homonuclear ones [39].

(4) Bonds are established in descending order of bond enthalpy (where positive values are assigned) until all formal valences of atoms are fulfilled, resulting in a total bond count per atom equating to $<Z> = 2$.

(5) The bond enthalpies of heteronuclear bonds can be correlated to those of homonuclear bonds via Pauling's equation [3]:

$$H_{XY} = \left[\left(\frac{H_{XX} + H_{YY}}{2}\right)\right] + [96.14(\eta_X - \eta_Y)^2] ----(12)$$

In the context of Pauling's equation, where $H_{XX}$ and $H_{YY}$ represent homonuclear bond enthalpies, $H_{XY}$ signifies heteronuclear bond enthalpy (measured in kJ/mol), while $Z_X$ and $Z_Y$ denote the electronegativities of atoms X and Y. For group IB–IIIB elements, Pauling's original data didn't encompass covalent bond enthalpies. A comparative analysis of Se–Se (330.5 kJ mol$^{-1}$) and Te–Te (293.3 kJ mol$^{-1}$) bonds indicates the sturdier nature of Se–Se bonds. This resilience aligns with Se's higher electronegativity relative to Te. As a consequence, Se–Zn bonds exhibit larger strength than Te–Zn bonds, influenced by the higher electronegativity of Se. This dynamic prompts Zn preference for Se within the parent glass network, causing a decrease in Se–Se bonds (330.5 kJ mol$^{-1}$) due to Zn integration into polymeric Se chains. This integration results in the formation of Se–Zn bonds (170.7 kJ mol$^{-1}$) and Zn–Zn bonds (22 kJ mol$^{-1}$). The cumulative outcome is a reduction in the system's cohesive energy as the concentration of Zn increases within the parent glass. This phenomenon likely contributes to the overall decrease in the mean bond energy and average heat of atomization [41, 42]. The decline in the mean bond energy and heat of atomization of the glass manifests in a corresponding decrease in the micro-hardness of these glass compositions [18,47].

Figures 12 and 13 distinctly illustrate the escalating microvoid volume ($V_h$) in correlation with an augmented zinc concentration within the parent STS glassy system. Each of these figures presents a distinct set of graphs. In Figure 12, the time interval spans from 5 seconds to 25 seconds, while the load remains constant, ranging from 100 gm to 300 gm. Remarkably, this configuration reveals a substantial surge in Vickers hardness for the materials, transitioning from 190 ± 20 (Å$^3$) to 270 ± 20 (Å$^3$) under a constant load subjected to varying time intervals. Figure 13, on the other hand, investigates load variation across the range of 100gm to 300gm, while maintaining a consistent time frame of 5 sec to 25 sec. Interestingly, the outcomes closely resemble those observed in Figure 12, providing a consistent pattern. In conclusion, the data points to a consistent behavior within our materials, irrespective of whether load or time is the variable under consideration. This coherence in outcomes lends strength to the conclusion that the system's response remains uniform and predictable in the face of alterations in load or time. In Figure 14(a), a distinct pattern emerges, illustrating the reciprocal connection between microhardness and micro void volume concerning coordination number. As the concentration of zinc rises within the parent sample, hardness demonstrates a decrease alongside the coordination number. Conversely, the micro void volume exhibits an elevation in tandem with the coordination number in the presence of increased zinc concentration. Figure 14(b) presents the correlation between the average coordination number <Z> and the compositional variations of both $T_g$ and $H_V$. Within this depiction, the diminishing trajectories of $T_g$ and $H_v$ are evident, highlighting the shared downward trends.

In Figure 15 (a), the interplay between the average coordination number and the compositional influence on the modulus of elasticity (*E*) is visually demonstrated. The graph reveals a noteworthy trend: the introduction of Zn leads to a decrease in the value of E. Significantly, the peak value materializes at Zn incorporation of 4 at%. This observation underscores the distinct impact of Zn on the mechanical properties, echoing the nuanced nature of its interaction with the glass matrix [43].

The fragility index plays a pivotal role in characterizing the glass transition behavior of materials. It offers a quantitative measure of how rapidly a material transitions from a glassy, solid-like state to a more fluid-like state as temperature increases. In essence, it gauges the sensitivity of a material's viscosity to changes in temperature. A high fragility index signifies that a material undergoes a sharp transition from a solid to a liquid state, exhibiting a rapid increase in viscosity as temperature rises. On the other hand, a low fragility index indicates a more gradual change in viscosity and a smoother transition from solid to liquid. This index is particularly relevant in various fields, including material science and industrial applications, as it provides insights into the behavior and usability of materials across different temperature ranges. The concept of the fragility index has proven valuable in predicting a material's response to changes in temperature and its suitability for specific applications. By analyzing the relationship between the heating rate *β* and the glass transition temperature $T_g$ suggested by Moynihan's theory of structural relaxation, we can calculate the corresponding activation $E_g$ using the following formula [49].

$$\ln(\beta) = \left(-\frac{E_g}{RT_g}\right) + constant ----- (13)$$

Following Equation (11), when plotting the natural logarithm of the heating rate $\ln(\beta)$ against the reciprocal of the glass transition temperature (1000/$T_g$), the resulting graph should exhibit a linear relationship. From the gradient of this linear graph, we can extract the $E_g$ linked to the molecular shifts and reconfigurations taking place near $T_g$. These plots were produced for both the original (ternary) and quaternary STSZ samples. The fragility index (*m*) can be computed using the following relation [50]:

$$m = \frac{E_g}{2.303RT_g} ----- (14)$$

The depicted Figure 15(b) clearly illustrates a notable trend: a decrease in the fragility index of the glass as the heating rate increases. This behavior is observed in both the parent glass sample and the quaternary STSZ alloy. The value of *m* signifies how rapidly a glass transitions from a supercooled liquid to a crystalline solid as the temperature rises. A higher fragility index suggests that the glass undergoes this transition more abruptly, while a lower fragility index implies a smoother and more gradual transition.

Now we come to another compositional aspect of the present samples through their mean coordination number <Z>. This parameter, extensively utilized in the characterization of network glass structures [44-47], remains agnostic to the specific atomic species involved in covalent bonding [48]. The number <Z> signifies the average count of neighboring atoms bonded to a central atom, serving as an indicator of the structural units' robustness [48]. The computation of <Z> for the examined glassy alloys adheres to the established methodology outlined by Tanaka [48], with coordination numbers (CN) of 4, 4, 2, and 2 attributed to Zn, Sn, Te, and Se, respectively. This approach offers a comprehensive and unified perspective on the composition-driven variations in the glassy alloy's structural arrangement.

$$<Z> = \frac{(eZ_{Se} + fZ_{Te} + gZ_{Sn} + hZ_m)}{(e + f + g + h)} ----- (15)$$

Where $Z_{Se}$, $Z_{Te}$, $Z_{Sn}$, and $Z_m$ are the coordination number of Se, Te, Sn, and metallic additive Zn respectively in $Se_{78-x}Te_{20}Sn_2Zn_x$ (0 ≤ x ≤ 6) glassy alloy.

In recent times, Phillips [44] introduced a novel perspective, interpreting the attributes of covalent glasses through the lens of <Z>. To elucidate the glass-forming capabilities exhibited by specific covalent glassy alloys, Phillips artfully employed constraint-counting arguments. This led him to propose that the connection of the network may have been elegantly designated by one variable <Z>. His analysis extended further by assuming that in the realm of an ideal glass (i.e., the interaction in three-dimensional space between the number of interatomic force-field restrictions per atom $N_c$ and the number of degrees of freedom per atom, $N_d$) follows a delicate equilibrium. This balance, encapsulated in Phillips' condition $N_c = N_d$, accommodates considerations encompassing the bending and stretching modes of the bonding constraints, ultimately culminating in the optimization of mechanical steadiness for a network possessing a critical coordination number <Z> = 2.12 Subsequently, Thorpe [46-48] recognized the intrinsic connection between Phillips' balance condition and the concept of percolation, where the focus shifts to the pivotal attribute of rigidity. Thorpe demonstrated that, in the context of a three-dimensional network, the count of zero-frequency (floppy) modes per atom, represented as f, conforms to the expression:

$$f = 2 - \frac{5}{6}<Z> -----(16)$$

Thorpe's insight lends a comprehensive framework, aligning Phillips' balance condition with the intricate dynamics of rigidity and zero-frequency modes, thus enriching our understanding of covalent glass behavior.

The lone pair of electrons is calculated by the given below formula

$$LP = V - <Z> ---(17)$$

Here $P$ denotes the total number of lone-pair electrons while $V$ is number of the unshared lone pair electrons [49, 50]. From Figure 16(a), The decrease in the degree of freedom and the count of lone pairs with increasing composition and coordination number can be attributed to the changing atomic structure and bonding environment within the glassy alloy. As the composition and coordination number increase, it indicates a higher degree of atomic ordering and connectivity within the material. In a glassy alloy, the degree of freedom refers to the number of independent ways in which atoms or particles can move without violating any constraints. As the coordination number increases, atoms become more closely packed and are constrained by stronger bonds and interactions. This reduced freedom of movement is reflected in the decrease in the degree of freedom. Higher coordination numbers result in a more rigid atomic arrangement, limiting the ability of atoms to move independently and freely. Lone pairs of electrons are non-bonding pairs of electrons that reside on atoms in the material. In glassy alloys, these lone pairs can influence the overall structural arrangement and chemical properties. As the composition and coordination number increase, the arrangement of atoms becomes more ordered and the bonds between them become stronger. This increased bonding reduces the availability of electrons for forming lone pairs. Additionally, higher coordination numbers often involve atoms being engaged in multiple bonds, leaving fewer lone pairs available. Consequently, the count of lone pairs decreases as the material becomes more densely packed and ordered.

Certainly, let's shift our focus to the experimental density. Density is a fundamental physical property that quantifies the mass of a substance per unit volume. In the context of the research you've mentioned, experimental density measurements were conducted to gain insights into the structural characteristics and packing arrangement of the glassy STSZ alloys. Experimental density values allow researchers to draw connections between the composition of the alloys and their structural properties. The density measurements for the current glassy alloys are carried out experimentally using the Archimedes method. We have utilized distilled water as a reference and executed measurements under room temperature conditions. The density can be calculated by employing the subsequent relation [51]:

$$d = \left(\frac{W_a}{W_a - W_l}\right) d_l \quad ----- (18)$$

In this context, the parameters $W_a$, $W_l$, and $d_l$ represent the sample's weight in an air environment, the sample's weight in the liquid, and the density of the chosen reference liquid. Density is a valuable physical parameter that provides insights into the properties of glassy materials, particularly amorphous materials. In glassy materials, density is closely connected to the average coordination number, which characterizes the arrangement of atoms in the material. By employing the following formula, the density of chalcogenide glasses can be determined [52-54]:

$$\rho = \left(\Sigma \frac{m_i}{d_i}\right)^{-1} \quad --- (19)$$

the fraction of mass ($m_i$) and the density ($d_i$) associated with each specific structural unit within the material. Considering the density equation mentioned earlier, we can derive the relationship between molar volume by expressing it as follows [52-54]:

$$V_m = \left(\frac{\Sigma x_i M_i}{\rho}\right) \quad --- (20)$$

The relationship between molar volume and density can be expressed by incorporating the atomic fraction of each component, represented by $x_i$, and their corresponding atomic mass, represented by $M_i$. The changes in density and molar volume with increasing composition and coordination number are shown in Figure 16(b) as an indication of the alterations in the atomic packing and bonding within the glassy alloy. These changes can be explained as follows:

Density is a measure of mass per unit volume. When the composition and coordination number increase in a glassy alloy, it generally implies a higher level of atomic ordering and denser packing. This increased atomic ordering leads to a more efficient use of space within the material. As atoms are arranged more closely and compactly, the mass per unit volume, i.e., the density, increases. The stronger bonds and reduced void spaces between atoms contribute to the higher density observed in materials with higher coordination numbers.

As the composition and coordination number increase, the arrangement of atoms becomes more ordered and closely packed. This results in a reduction in the amount of free space available for atoms to occupy. Consequently, the volume filled by one mole of the substance, i.e., the molar volume, decreases. The increase in atomic bonding and reduction in void spaces cause the material to take up less volume per mole, leading to the observed inverse relation between molar volume and coordination number [52-54].

## 5. Conclusion

The exploration of glassy STSZ alloys has unveiled insights into their compositional influence on thermal stability, microhardness, and compactness. The fluctuations in micro-hardness concerning composition have been thoroughly scrutinized, and the impact of Zn integration on the micro-hardness of these glasses finds an enlightening explanation rooted in the average bond energy and average heat of atomization within the glass matrix. Remarkably, the results underscore that the entire spectrum of thermo-mechanical parameters investigated in this study achieves their most extreme values at a Zn incorporation of 4 at%. However, amidst these intricate variations, a notable observation emerges: the energy of micro-void formation within the glassy network of quaternary glasses remains relatively unchanged after the introduction of Zn into the parent STS alloy. This distinctive detail underscores the nuanced interplay of factors at play, signaling a balanced equilibrium within the glassy structure even amid compositional alterations.

**CRediT authorship contribution statement**

Vishnu Saraswat and N. Mehta: Formal analysis, Plotting graphs, Writing-original draft

A. Dahshan and H. I. Elsaeedy: Conceptualization, Revising-final draft.

**Acknowledgments:** The authors extend their appreciation to the Deanship of Scientific Research at King Khalid University for funding this work through large group Research Project under grant number RGP.2/158/44.

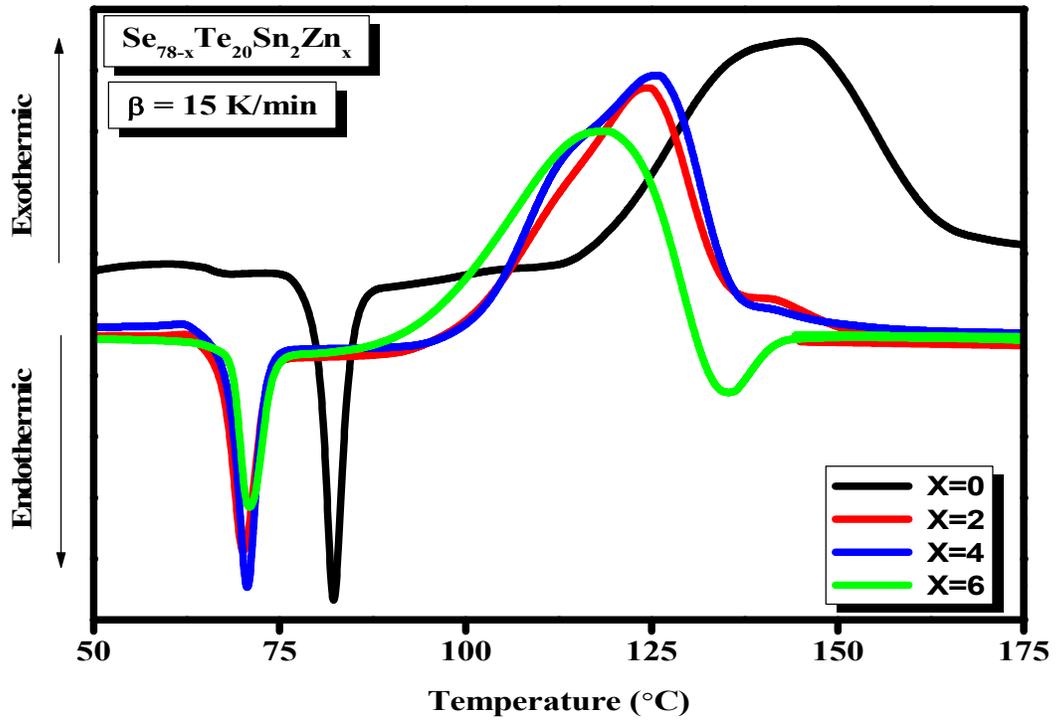

Figure1 The presence of distinct glass transition peaks in the DSC scans was observed for all samples, confirming their inherent nature.

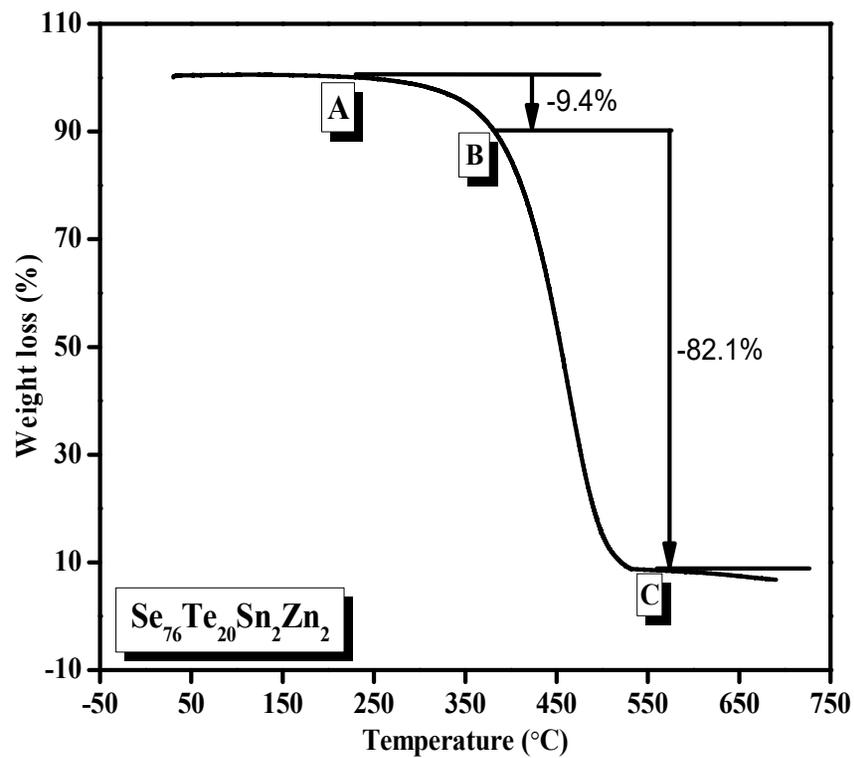
(a)

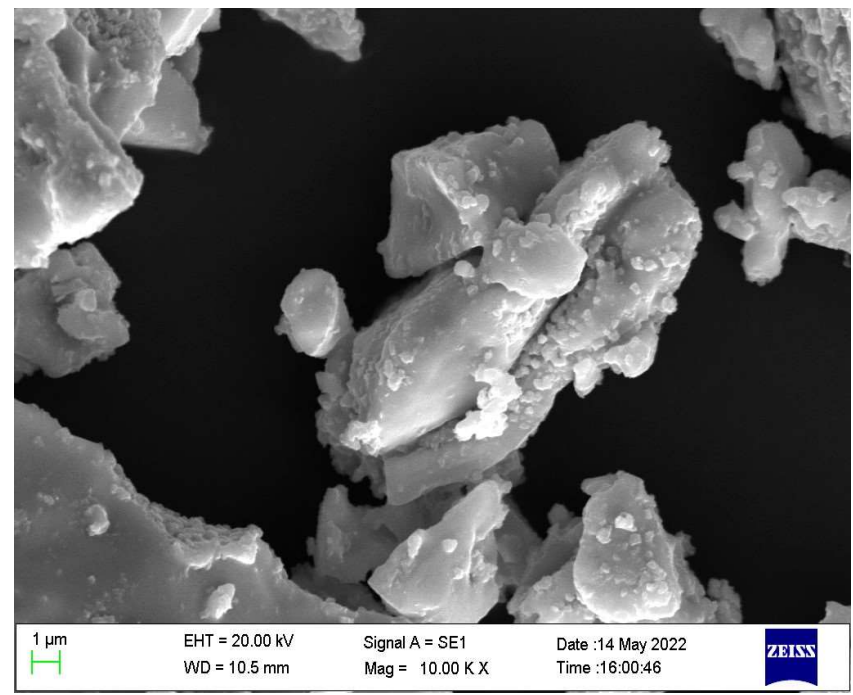
(b)

Figure 2(a) Thermogravimetric analysis (TGA) profile of the glassy Se$_{78-x}$Te$_{20}$Sn$_2$Zn$_x$ (x=2) sample. (b)An SEM image of a representative sample Se$_{78-x}$Te$_{20}$Sn$_2$Zn$_x$ (x = 2) was obtained to verify its glassy nature

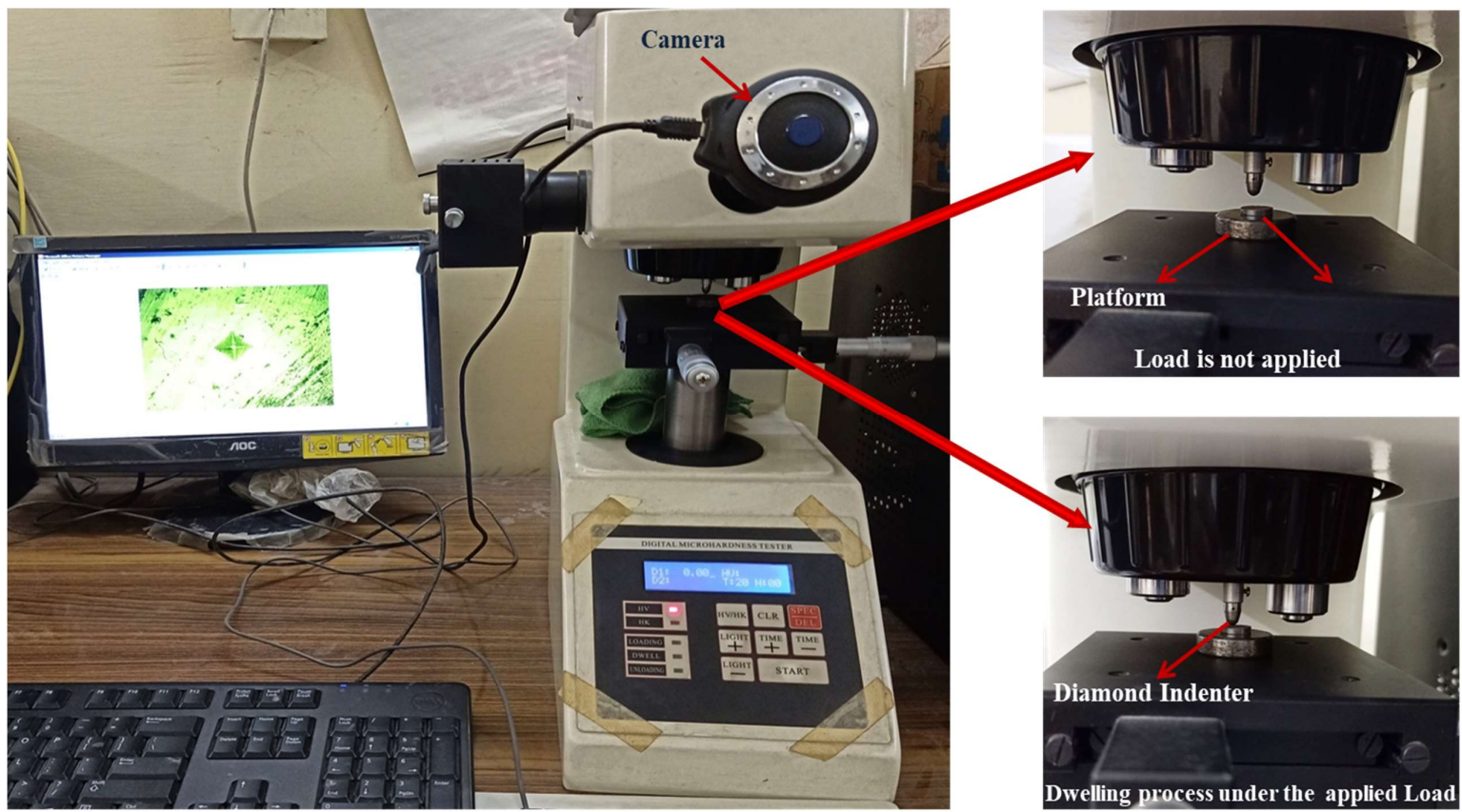

Figure 3 Image of the microscopic setup outfitted with the micro-indentation capability.

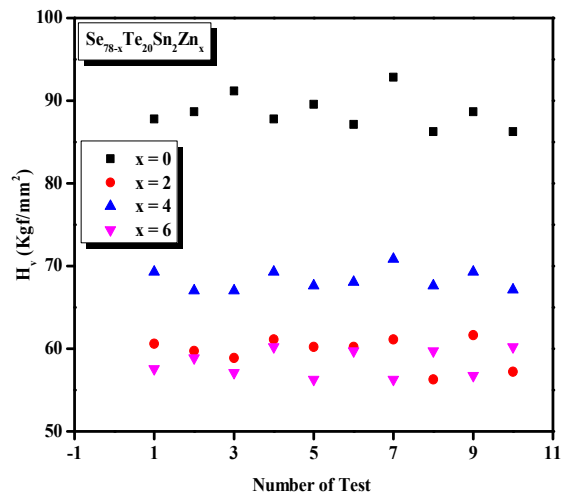 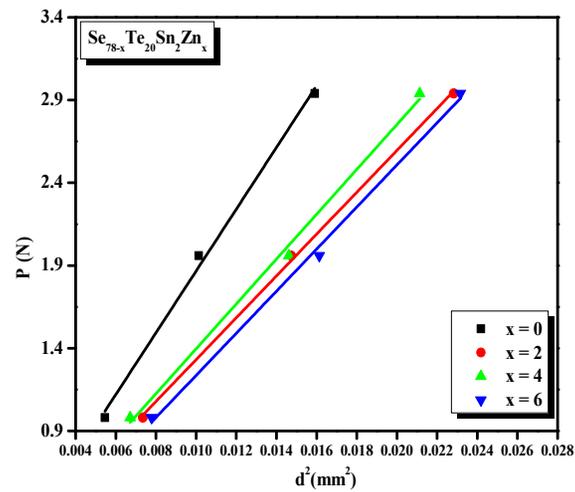 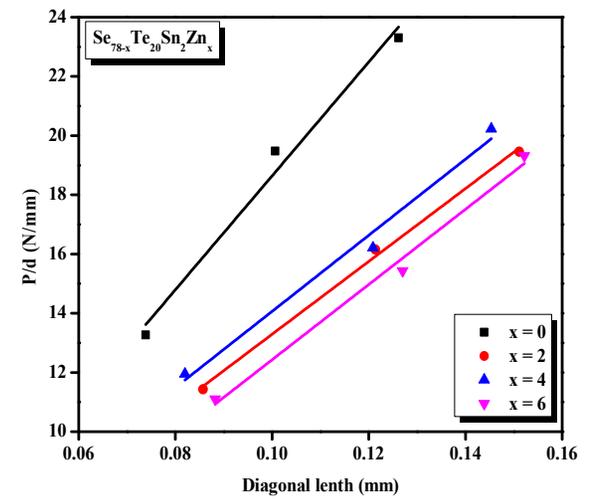

(a) (b) (c)

Figure 4 (a) Microhardness measurements were conducted across multiple iterations. (b) The graph depicts the correlation between the applied load and the diagonal square linear fitting for the glassy $Se_{78-x}Te_{20}Sn_2Zn_x$ samples, where the parameter x takes on values from 0 to 6.

(c) The linear fitting of P/d vs diagonal length is illustrated for the glassy $Se_{78-x}Te_{20}Sn_2Zn_x$ samples, where x ranges from 0 to 6.

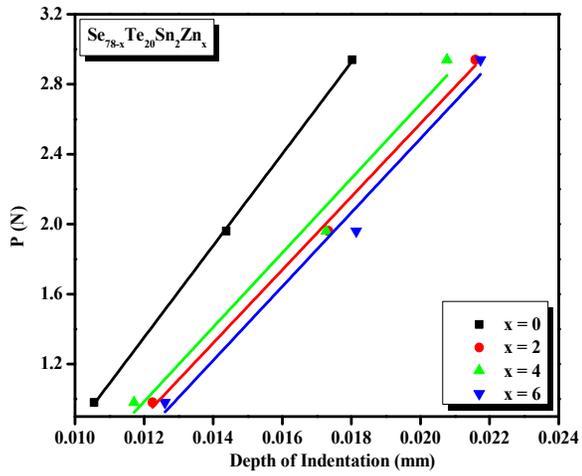
(a)

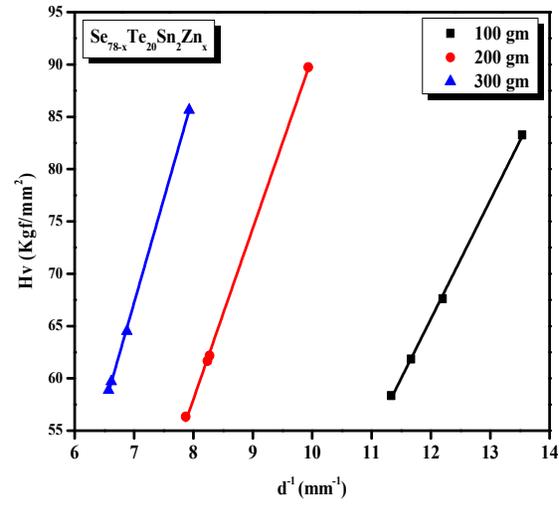
(b)

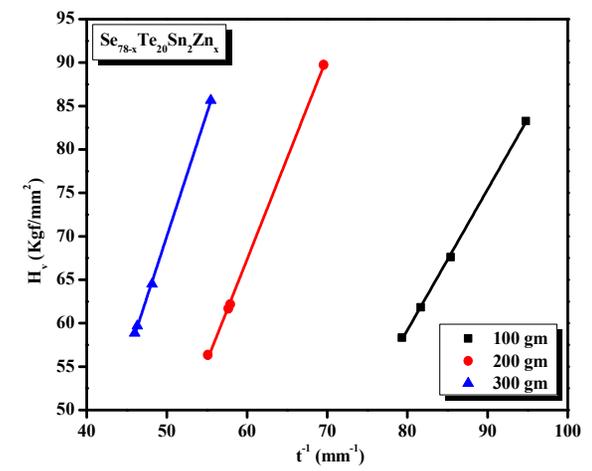
(c)

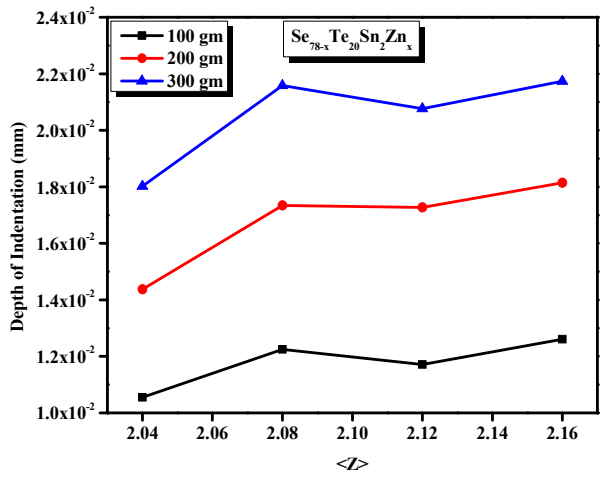
(d)

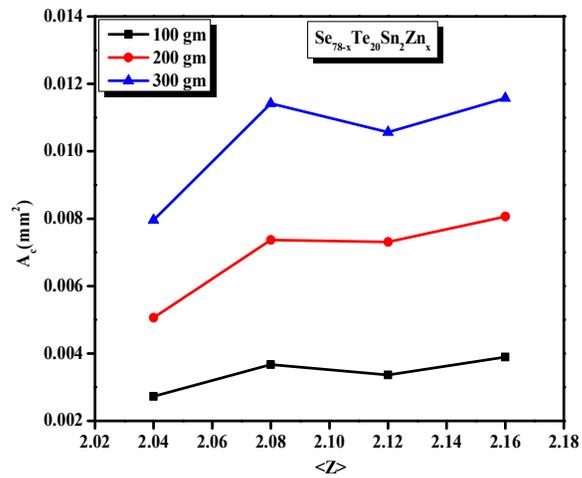
(e)

Fig 5 (a)The graph illustrates the relationship between the applied load and the depth of indentation for the glassy $Se_{78-x}Te_{20}Sn_2Zn_x$ samples, where the variable x ranges from 0 to 6. (b) The linear fitting of $H_v$ vs the inverse of diagonal length is conducted for the glassy $Se_{78-x}Te_{20}Sn_2Zn_x$ samples, covering the range of x values from 0 to 6. This analysis is performed under constant loads of 100 gm, 200 gm, and 300 gm. (c) A linear fitting of $H_v$ against the inverse of depth of indentation is conducted for the glassy $Se_{78-x}Te_{20}Sn_2Zn_x$ samples, encompassing the parameter x ranging from 0 to 6. This analysis is carried out under constant loads of 100 gm, 200 gm, and 300 gm.(d) A study of the relationship between the depth of indentation and coordination number is conducted for the glassy $Se_{78-x}Te_{20}Sn_2Zn_x$ samples, covering the parameter x within the range of 0 to 6 and under consistent loads of 100 gm, 200 gm, and 300 gm. (e) An exploration into the correlation between the Vickers contact area and coordination number is undertaken for the glassy $Se_{78-x}Te_{20}Sn_2Zn_x$ samples. This investigation encompasses a range of x values from 0 to 6 and is conducted under uniform loads of 100 gm, 200 gm, and 300 gm

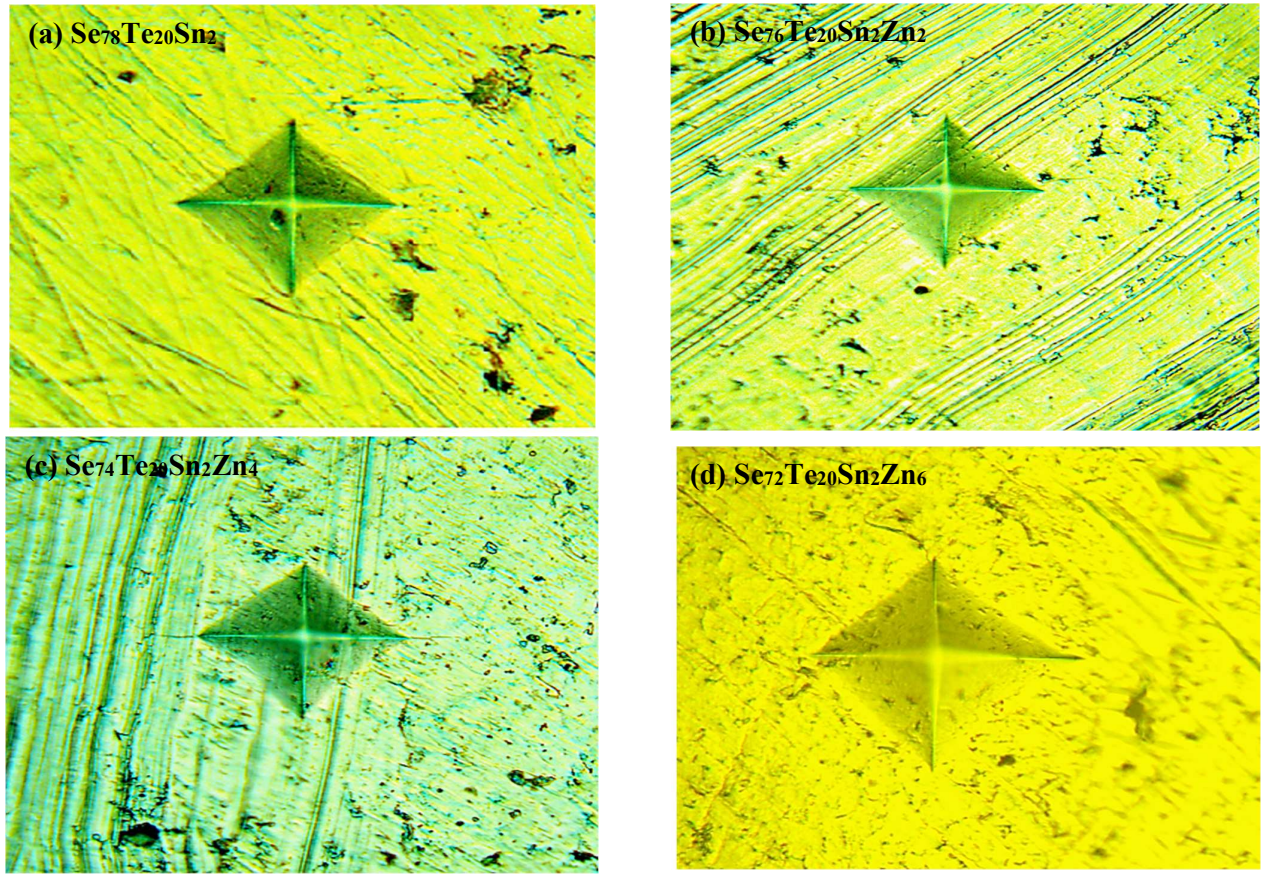

Figure 6 Micrographs were taken of the Vickers indents on the surfaces of bulk amorphous alloys with compositions $Se_{78-x}Te_{20}Sn_2Zn_x$, where x represents values of 0, 2, 4, and 6 at seconds and 100 gm loads.

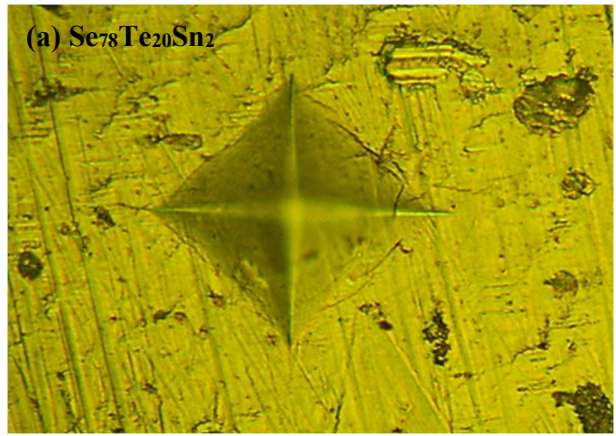 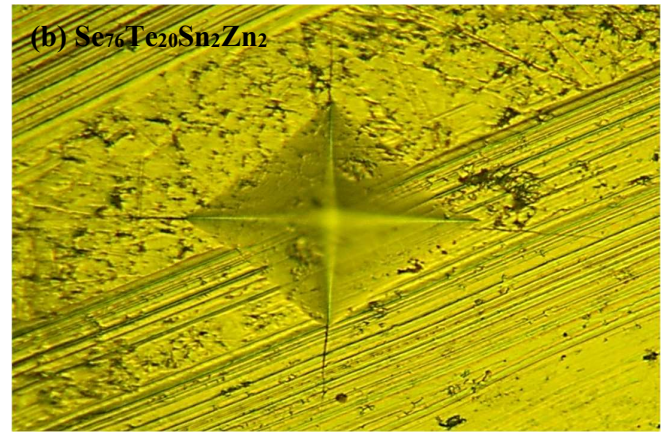
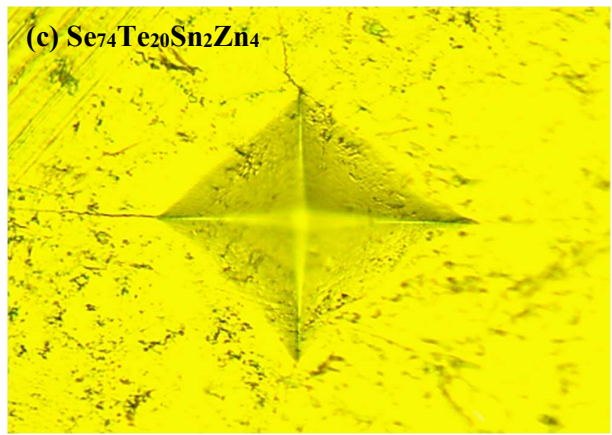 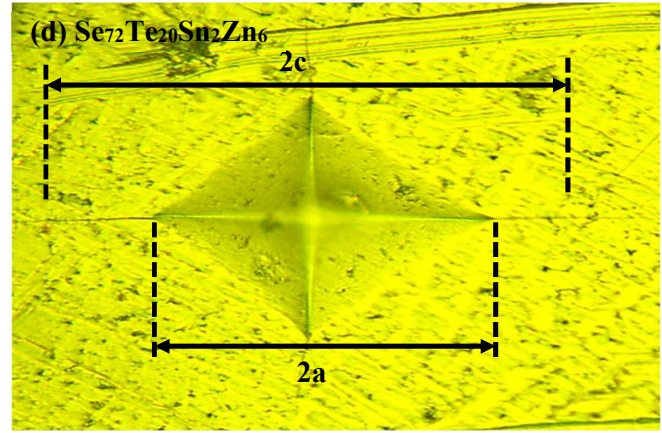

Figure 7 Micrographs were taken of the Vickers indents on the surfaces of bulk amorphous alloys with compositions $Se_{78-x}Te_{20}Sn_2Zn_x$, where x represents values of 0, 2, 4, and 6 at 25 seconds and 200 gm loads.

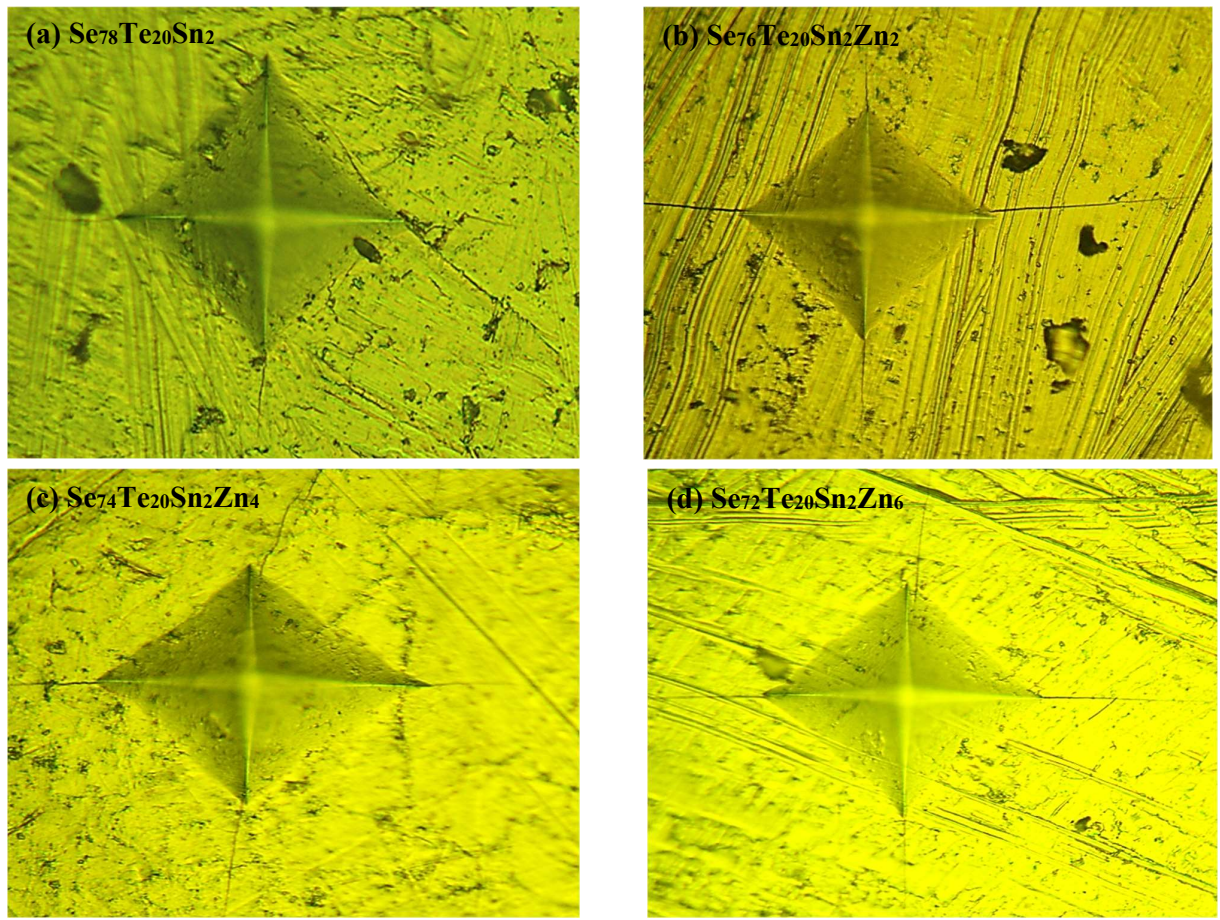

Figure 8 Micrographs were taken of the Vickers indents on the surfaces of bulk amorphous alloys with compositions $Se_{78-x}Te_{20}Sn_2Zn_x$, where x represents values of 0, 2, 4, and 6 at 25 seconds and 300 gm loads.

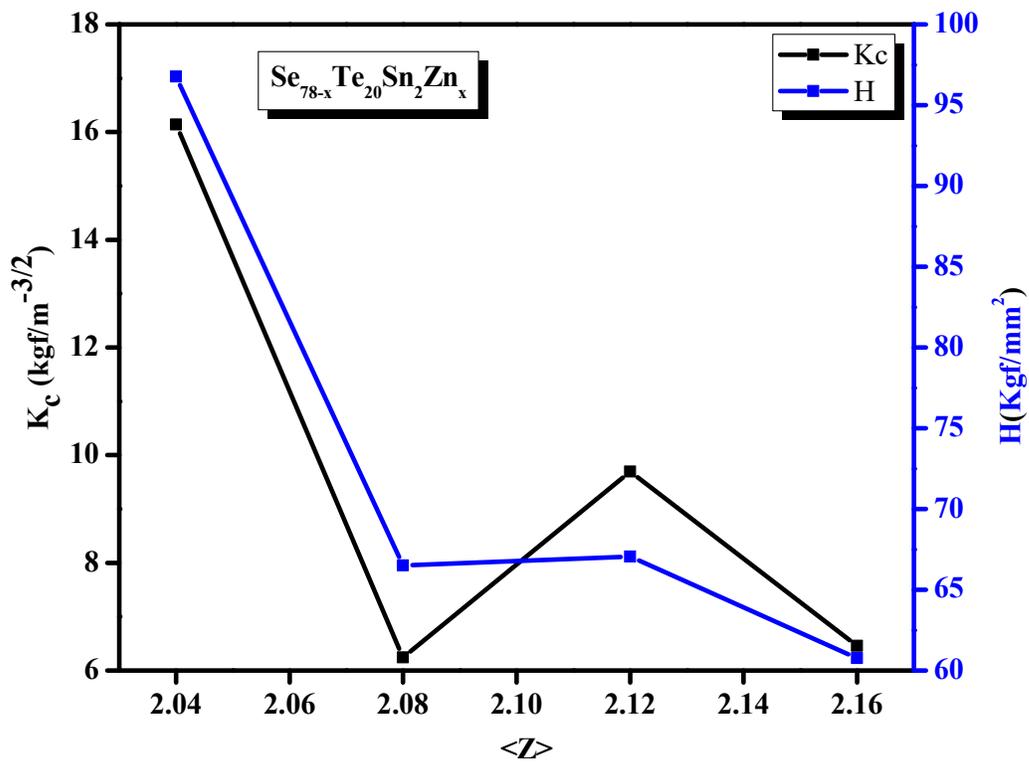

Figure 9 Indentation toughness ($K_c$) and hardness ($H$) of STSZ glasses as a function of the mean coordination number.

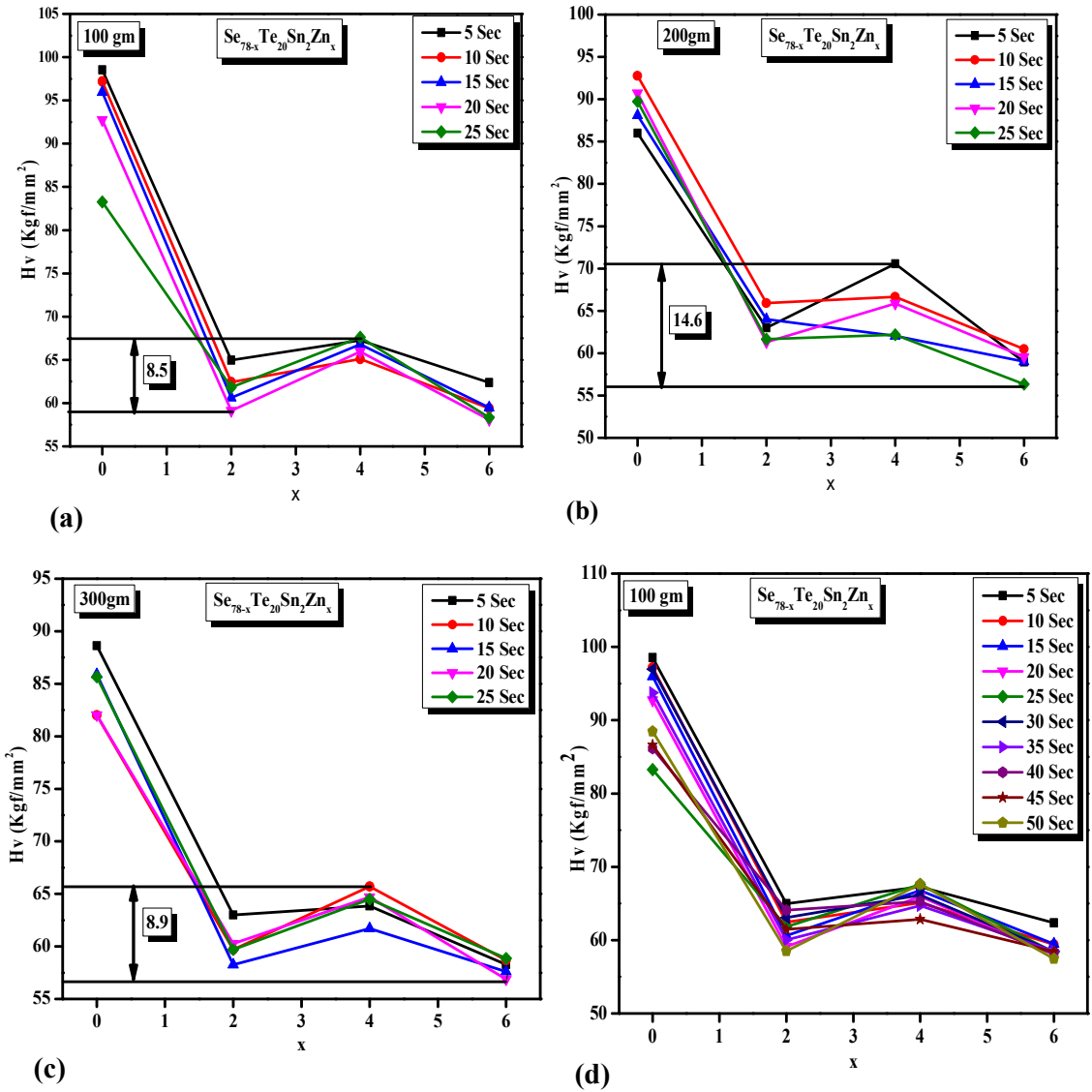

Figure 10 Vickers hardness ($H_v$) against the composition for $Se_{78-x}Te_{20}Sn_2Zn_x$ (x = 0, 2, 4, 6) amorphous alloys at constant load 100 gm, 200 gm and 300 gm at different times.

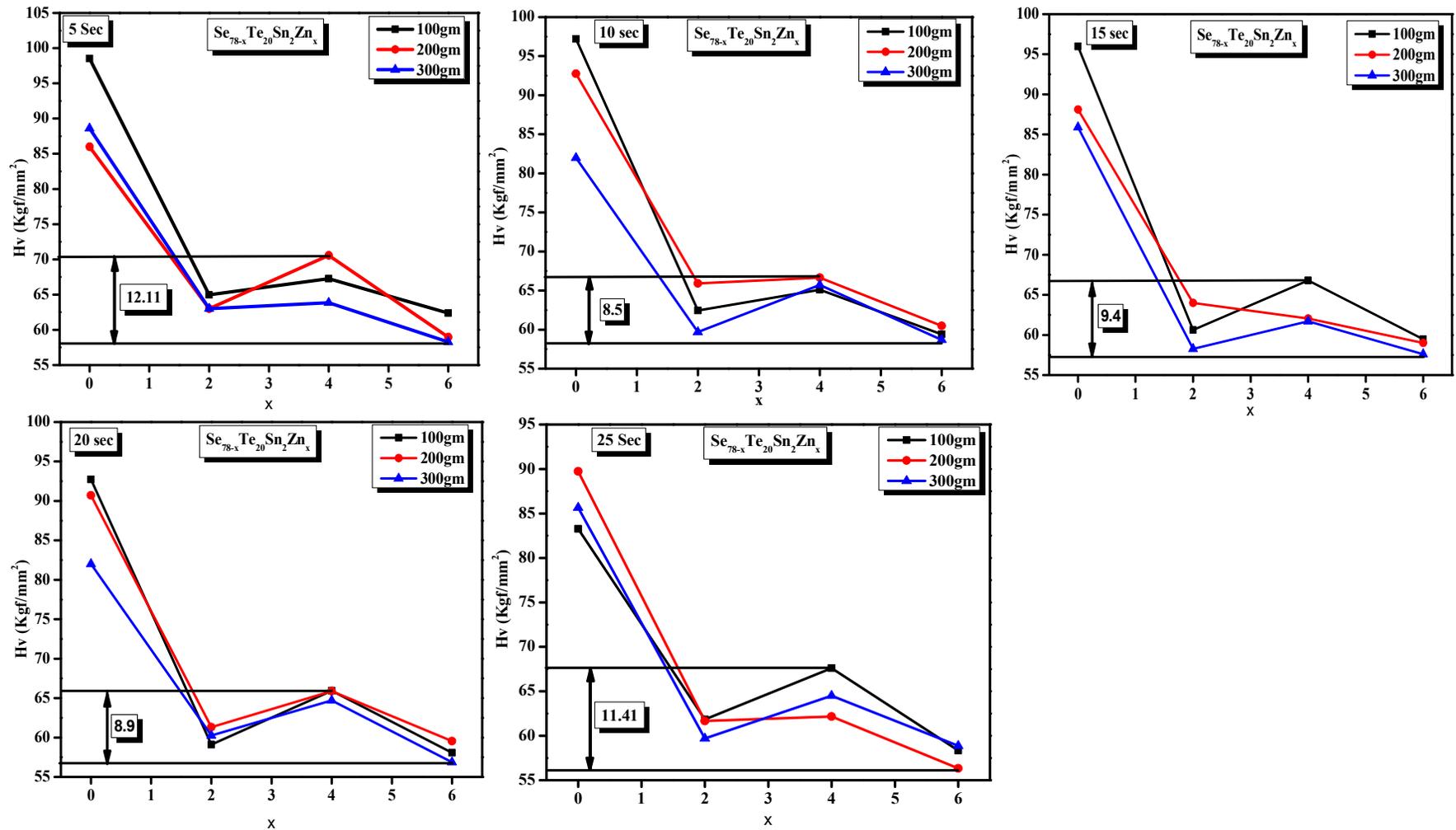

Figure 11 Vickers hardness (Hv) against the composition for $Se_{78-x}Te_{20}Sn_2Zn_x$ (x = 0, 2, 4, 6) amorphous alloys at varying load 100gm, 200gm and 300gm at constant time

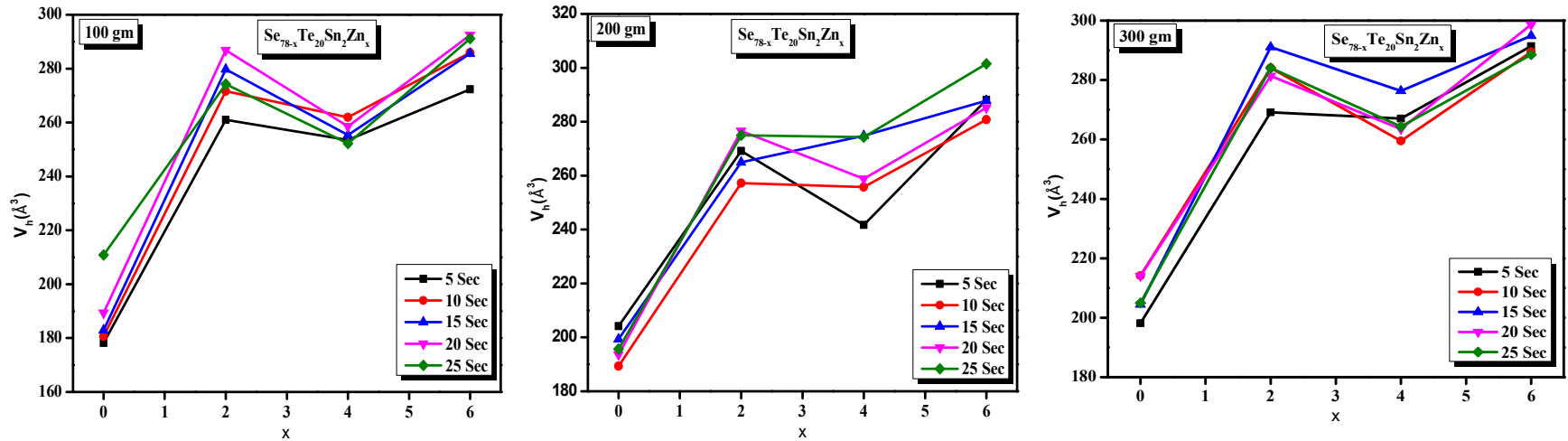

Figure 12 Micro-void volume ($V_h$) against the composition for $Se_{78-x}Te_{20}Sn_2Zn_x$ (x = 0, 2, 4, 6) amorphous alloys at constant load 100gm, 200gm and 300gm & varying time

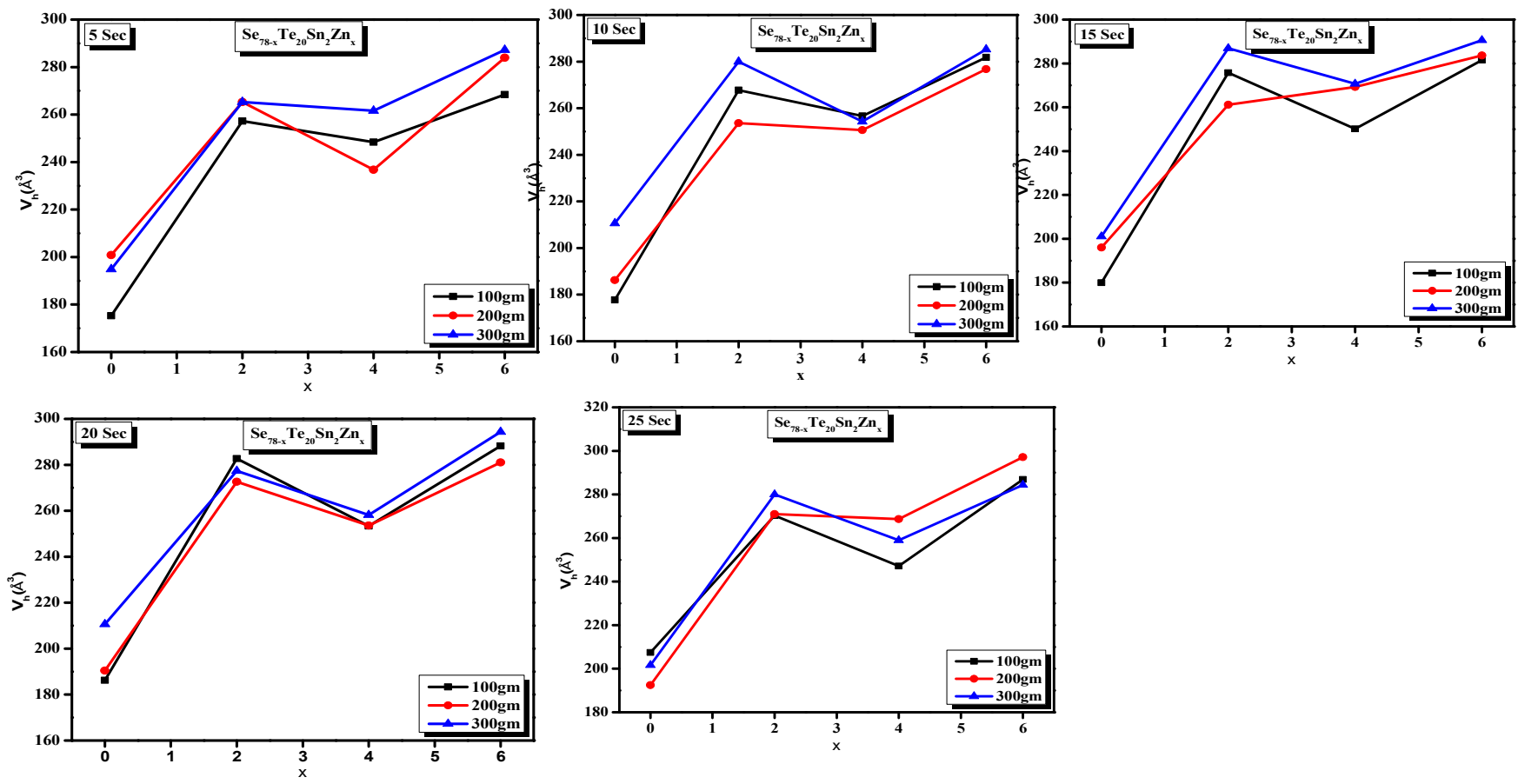

Figure 13 Vickers hardness ($H_v$) against the composition for $Se_{78-x}Te_{20}Sn_2Zn_x$ (x = 0, 2, 4, 6) amorphous alloys at varying time and constant load 100gm, 200gm and 300gm

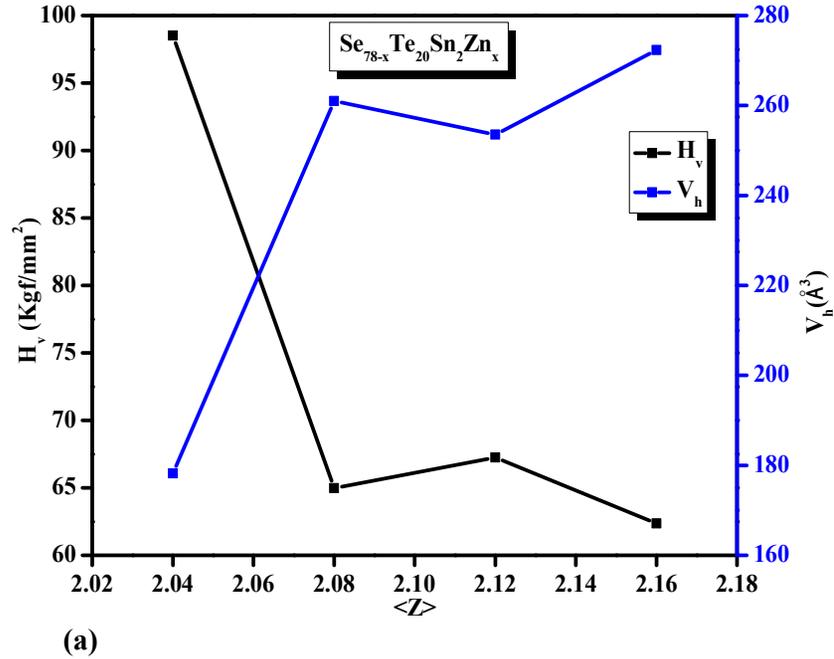 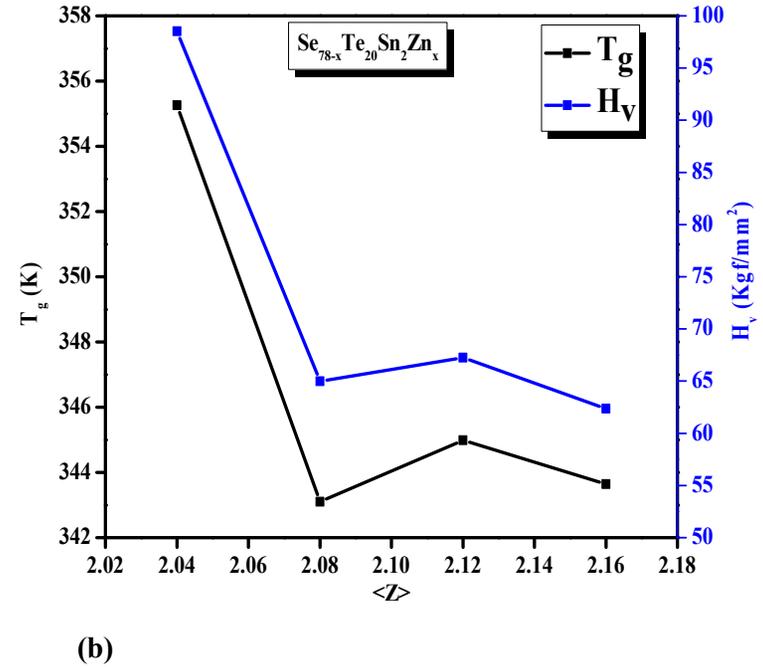

Figure 14 (a) The correlation between micro-void volume ($V_h$) and Vickers hardness ($H_v$) is investigated with respect to the mean coordination number in $Se_{78-x}Te_{20}Sn_2Zn_x$ (x = 0, 2, 4, 6) amorphous alloys. (b) The relationship between glass transition temperature ($T_g$) and Vickers hardness ($H_v$) concerning the average coordination number is explored for amorphous alloys with varying compositions of $Se_{78-x}Te_{20}Sn_2Zn_x$ (x = 0, 2, 4, 6).

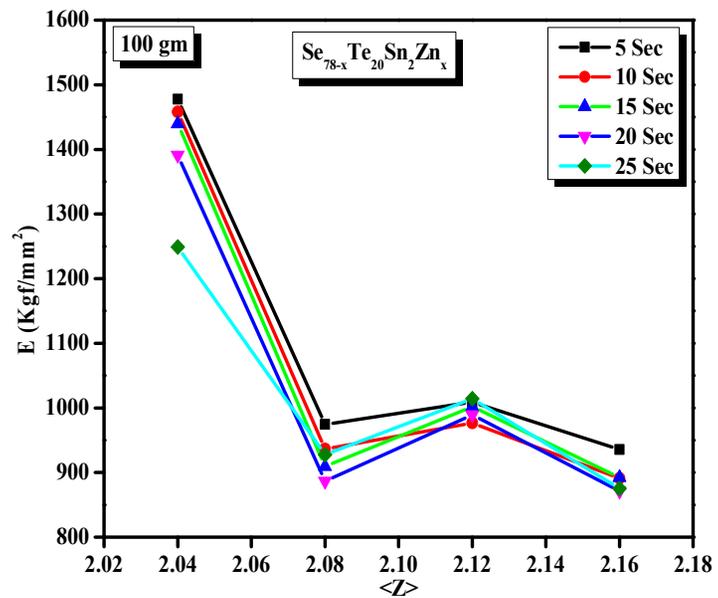 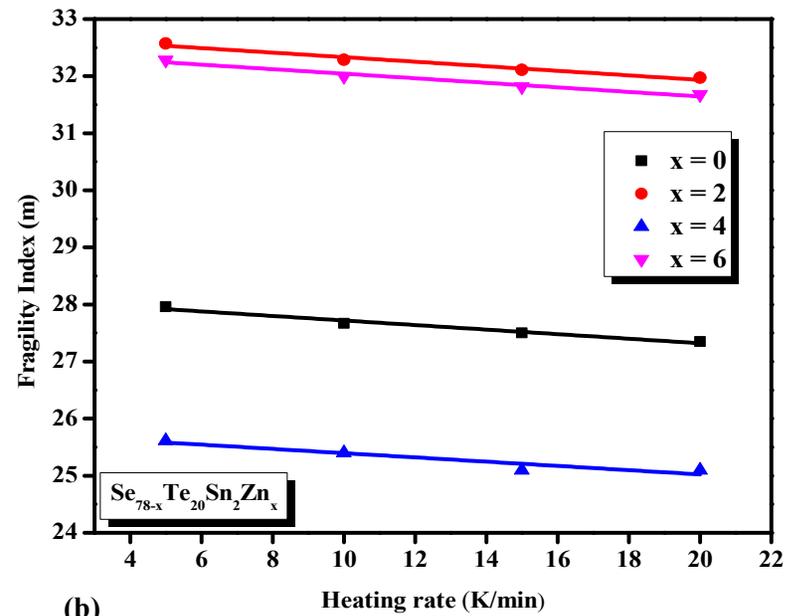

Figure 15 Plot of (a) the modulus of rigidity (*E*) against the average coordination number <Z>, and (b) the fragility index against the heating rate.

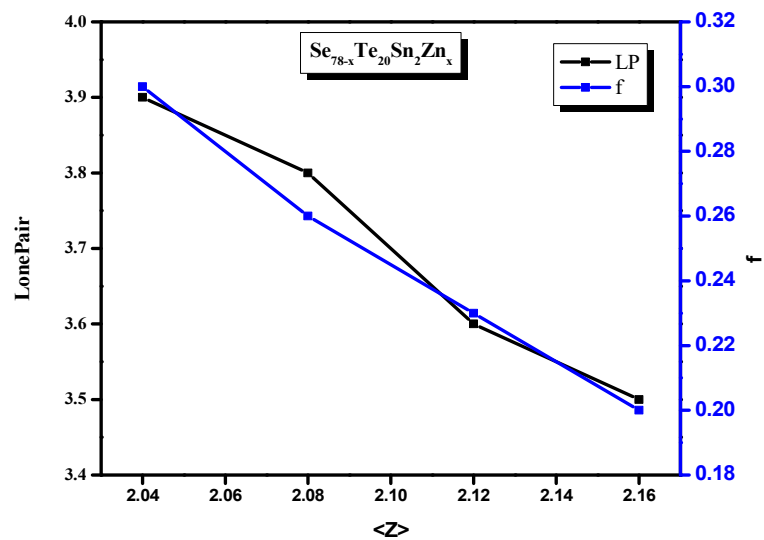 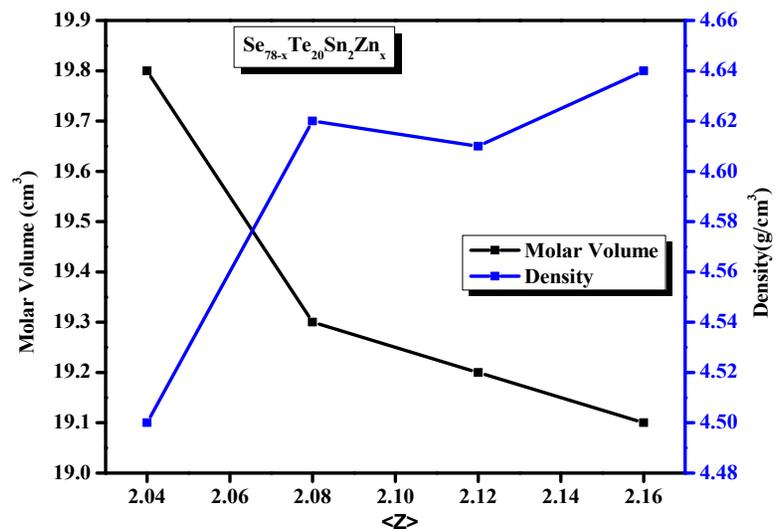

Figure 16 Plots of (a) lone pair and degree of freedom against average coordination number <Z>, and (b) molar volume and density against average coordination number <Z>.